\documentclass[structabstract]{aa} 
\usepackage{graphicx}
\usepackage{txfonts}
\begin{document}
   \title{The spectroscopic evolution of the recurrent nova T Pyxidis
during its 2011 outburst}
   \subtitle{II. The optically thin phase and the structure of the ejecta in recurrent novae}

   \author{S. N. Shore\inst{1,2}, G. J. Schwarz\inst{3}, I. De Gennaro Aquino\inst{1}, T. Augusteijn\inst{4},  F. M. Walter\inst{5}, S. Starrfield\inst{6}, \&   E. M. Sion\inst{7} }
        \institute{
 Dipartimento di Fisica ``Enrico Fermi'', Universit\`a di Pisa, and INFN- Sezione Pisa, largo
B. Pontecorvo 3, I-56127 Pisa, Italy; \email{shore@df.unipi.it,ivan.degennaroaquino@gmail.com}
\and
INFN- Sezione Pisa, largo B. Pontecorvo 3, I-56127 Pisa, Italy
  \and
  American Astronomical Society, 2000 Florida Ave NW, Washington DC 20009-1231, USA; \email{Greg.Schwarz@aas.org}
    \and
  Nordic Optical Telescope, Apartado 474, E-38700 Santa Cruz de La Palma,
Santa Cruz de Tenerife,Spain; \email{tau@not.iac.es}
 \and
 Department of Physics and Astronomy, Stony Brook University Stony Brook NY 11794-3800, USA; \email{frederick.walter@stonybrook.edu}
 \and
 School of Earth and Space Exploration,  Arizona State University, P.O. Box 871404, Tempe, AZ 85287-1404,USA  \email{sumner.starrfield@asu.edu} 
 \and
Department of Astronomy and Astrophysics, Villanova University, 800 Lancaster Avenue, Villanova, PA 19085, USA; \email{edward.sion@villanova.edu}
   }
     
              \date{received ---; accepted ---}

%DRAFT 3/9/12; after comments from Sumner

 \abstract
 {}
  % aims heading (mandatory)
   {We continue our study of the physical properties of the recurrent nova T Pyx, focussing on the structure of the ejecta in the nebular stage  of expansion during the 2011 outburst.}
   % methods heading (mandatory)
{ The nova was observed contemporaneously with the Nordic Optical Telescope (NOT), at high resolution spectroscopic resolution (R $\approx$65000) on 2011 Oct. 11 and 2012 Apr. 8 (without absolute flux calibration), and with the Space Telescope Imaging Spectrograph (STIS) aboard the Hubble Space Telescope, at high resolution (R$\approx$ 30000) on 2011 Oct. 10 and 2012 Mar. 28 (absolute fluxes).   The NOT spectra cover 3800-7300\AA, the HST spectra from 2011 Oct. cover 1150-5700\AA\ while the 2012 Mar. spectrum covers 1150-1700\AA.   We use standard plasma diagnostics (e.g. [O III] and [N II] line ratios and the H$\beta$ line fluxes) to constrain electron densities and temperatures.  Using Monte Carlo modeling of the ejecta, we derive the structure and filling factor from comparisons to the optical and ultraviolet line profiles. }
   % results heading (mandatory)
{The ejecta can be modeled using an  axisymmetric conical -- bipolar -- geometry with a low inclination of the axis to the line of sight, $i=15\pm5$ degrees, compatible with published results from high angular resolution optical spectro-interferometry.  The structure is similar to that observed in the other short orbital period recurrent novae (e.g. CI Aql, U Sco) and RNe candidate KT Eri during their nebular stages.  We show that the electron density scales as $t^{-3}$ as expected from a ballistically ejected  constant mass shell; there is no need to invoke a continuing mass outflow following the eruption.  The derived mass for the ejecta with filling factor $f \approx 3\%$, M$_{ej} \approx  2\times 10^{-6}$M$_\odot$ is similar to that obtained for other recurrent nova ejecta but inconsistent with the previously reported extended optically thick epoch of the explosion.  We suggest that the system underwent a common envelope phase following the explosion that produced the recombination event.  Implications for the dynamics of the recurrent novae are discussed. }
  % conclusions heading (optional), leave it empty if necessary 
  {The compact recurrent novae can be understood within a single phenomenological model with bipolar, although not  jet-like, low mass ejecta. }


          \thanks{Based on observations made with the NASA/ESA Hubble Space Telescope, obtained from the data archive at the Space Telescope Science Institute. STScI is operated by the Association of Universities for Research in Astronomy, Inc. under NASA contract NAS 5-26555.}
          \thanks{Based on observations made with the Nordic Optical Telescope, operated on the island of La Palma jointly by Denmark, Finland, Iceland, Norway, and Sweden, in the Spanish Observatorio del Roque de los Muchachos of the Instituto de Astrofisica de Canarias. }

  \titlerunning{The 2011 outburst of T Pyx. II. The Nebular Phase} \authorrunning{S. N.
Shore et al.}

  \maketitle

\section{Introduction}

Recurrent novae (RN)  occur in binary systems where mass accretion from the companion causes a 
thermonuclear runaway on a white dwarf (WD).  Unlike classical novae, originating in systems with very similar overall properties, the frequency of explosion and the low ejecta mass is taken to indicate that the WD is very close to the Chandrasekhar mass limit.  Since the explosion is initiated by the accumulation of a hydrogen layer, and its intensity is governed by the extent of deep envelope mixing with the accreted matter, such a massive WD requires a lower accumulated mass to reach ignition conditions (see e.g. Starrfield et al. 2008).  The source of the accreted material is less important than the nature of the mass gainer.  Accretion is  either from winds in a giant secondary in large separation systems (e.g. RS Oph, V407 Cyg, T CrB, periods $>$ several hundred days) or Roche lobe overflow of a late type, main sequence star or a more evolved but still low mass companion in short period systems (e.g. CI Aql, V394 CrA, U Sco, periods $<$ a few days).  With an orbital period of 1.83 hours (Uthlas et al. 2010), T Pyx is in the latter class of
RN and has the shortest orbital of any RN.  Mixed core and hydrogen rich accreted matter is violently ejected at
high velocities from a  few 10$^3$ to 10$^4$ km s$^{-1}$.  Any 
hydrogen not ejected from the WD continues to burn in hydrostatic
equilibrium until consumed or ejected via a wind or common envelope
mass ejection.  It is the extremely high mass of the WD that gives
RNe their characteristics of short recurrence timescales, low mass ejection 
($\sim$ 10$^{-6}$ M$_{\odot}$) at large velocities, and hence rapid evolution.

The long awaited {\it sixth} historical outburst of the recurrent nova (RN) T Pyx was detected
on 2011 April 14.2931 (MJD 55665.79310) by Waagan et al. (2011) .   While this was not the first outburst for which spectroscopy was available, it was the first of the era of observations with linear digital detectors coupled to high resolution spectrographs providing absolute calibrations, and for which contemporaneous panchromatic ground and space based observations from $\gamma$-ray through centimeter radio were possible.  In our first paper (Shore et al. 2011, hereafter Paper I) we presented high resolution optical spectra from the optically thick stage of the expansion,  determined a new and larger distance, $\ge 3.5$ kpc, and reddening, E(B-V)=0.5,  than previously thought,  and discussed the diagnostics of the dynamics and structure of the ejecta.  In this paper, we show the results from the late-time optical and ultraviolet high resolution observations following the transition to the nebular phase.   We show that  the line profiles from a range of ionizations and covering a very wide range of physical conditions provide a unified picture for the ejecta of T Pyx and, by extension, for all recurrent novae arising in compact binary systems.
%\footnote{The long quiescence has produced a variety of hypothesis on the ``real'' nature of this object, e.g. Schaefer, Pagnotta, \& Shara (2010) even %suggested that T Pyx is not a genuine RN but rather a CN whose explosion occurred in 1866 and now showing dwarf nova outburst.}  

\section{Observational Data}

Our optical data set, a continuation of the sequence described in Paper I, consists of spectra taken on 
2011 Oct. 11 (hereafter epoch 1) and 2012 Apr. 8 (hereafter epoch 2) with the 2.6 m Nordic Optical Telescope
(NOT) FIber-fed Echelle Spectrograph (FIES)  with a dispersion of 0.023 \AA\ px$^{-1}$ in
high-resolution mode, covering the spectral interval from 3635 to
7364 \AA\ and 0.035 \AA\ px$^{-1}$ in medium-resolution mode, covering
the spectral interval from 3680 to 7300\ AA\ (see Table 1a).    Both observations were done with the high-resolution fiber
The sequence was not absolutely calibrated. All NOT
spectra were reduced using IRAF, FIESTool, and IDL.
\footnote{IRAF is   distributed by the National Optical Astronomy Observatories, which
  are operated by the Association of Universities for Research in
  Astronomy, Inc., under cooperative agreement with the US National
  Science Foundation.}     Associated with these, we had two nearly simultaneous observing sequences  with the Space Telescope Imaging Spectrograph (STIS) aboard the {\it Hubble Space Telescope} (HST).   In the first, we obtained medium resolution echelle spectra from 1150-3100\AA\ and a single low resolution G430L optical spectrum from  2900-5700\AA.  Note that although the first epoch observation included spectra longward of 1700 \AA,  the epoch 2  data used the E140M grating between 1150 and 1700 \AA\ so we limit our discussion to the common spectral interval.  A more complete description of the full NOT and STIS data sets is in preparation.  The optical epoch 1 STIS spectrum was used for calibration of the nearly simultaneous NOT spectrum.   The epoch 2 NOT spectrum was absolutely calibrated with a low resolution SMARTS spectrum from 2012 Apr. 2.    The journal of STIS observations is given in Table 1b.  {\bf The dates relative to the discovery of the outburst are also listed.}
  %}
  
\begin{center}
Table 1a: Journal of NOT/FIES observations \\
\begin{tabular}{ccccr}
\hline
Date  & Time (UT) & MJD  & Day & t$_{exp}$ (sec)\\
\hline
2011 Oct. 11& 05:51 & 55845.744 & 180 & 600 \\
2012 Apr. 8 &  20:50 & 56026.368 & 360 & 3600 \\
\hline
\end{tabular}
\end{center}
{\small 
\begin{center}
Table 1b: Journal of {\it HST}/STIS observations \\
\begin{tabular}{ccrc}
\hline
OBSID &  MJD  & t$_{exp}$ (sec) & Grating \\
\hline
&  2011 Oct.  4  (GO 12200) &  Day 172 &  \\
\hline
obg103010 &  55838.540 & 600 & E140M \\
obg103020 &  55838.552 & 600 & E230M  \\
obg103040 &  55838.567  & 155 & E230M  \\
obg103030 &  55839.061 & 5 & G430L \\
\hline
 & 2012 Mar. 28  (GO/DDT 12700) & Day 349 & \\
 \hline
obx701010 &  56015.142 & 2457 & E140M  \\
obx701020 & 56015.201 & 3023 & E140M  \\
\hline
\end{tabular}
\end{center}
}

\section{Analysis}
\begin{figure*}
   \centering
   \includegraphics[width=16cm]{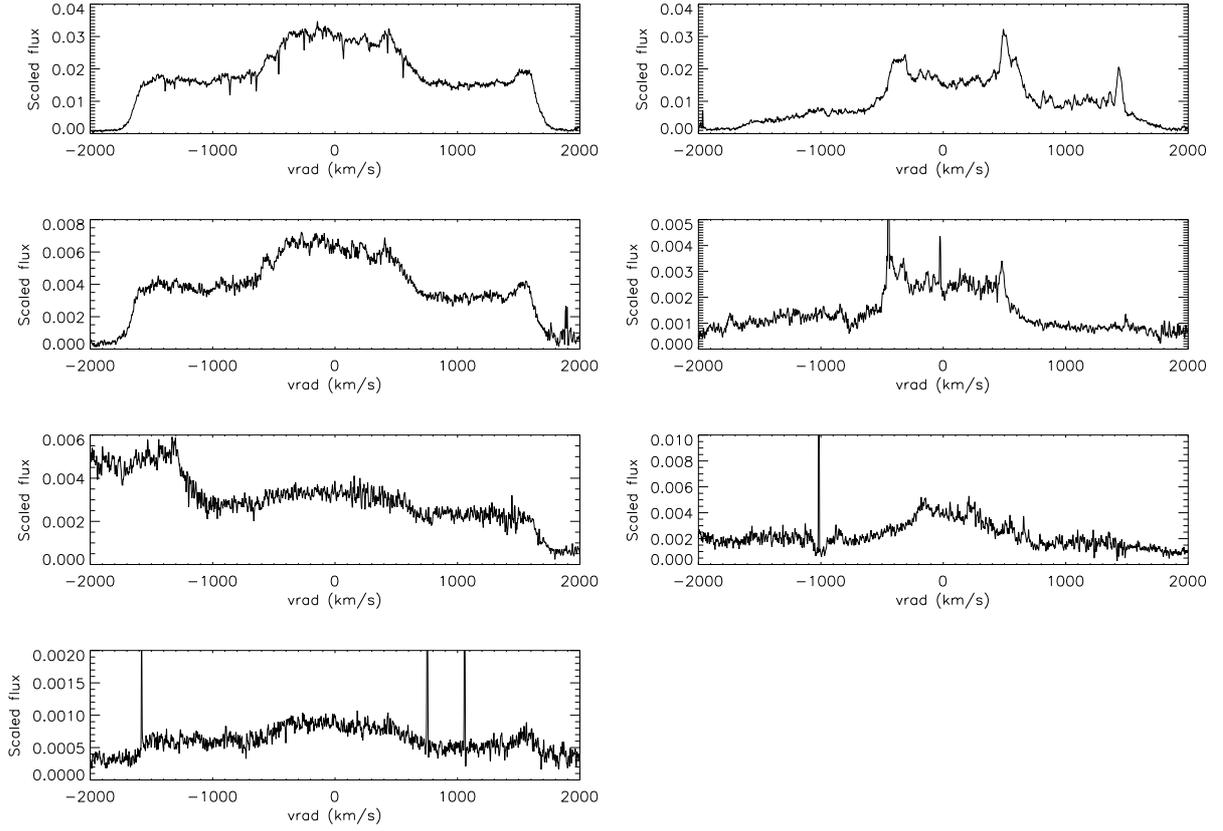}
   \caption{H$\alpha$ (first row), H$\beta$ (second row), He II 4686\AA\ (third row), and He I 5875\AA\ (bottom, left) line profiles in radial velocity (km s$^{-1}$, uncorrected for LSR) epoch 1 (left)  and epoch 2 (right), from the  NOT spectra.   Note that He I was not present in the epoch 2 spectra. }
    \end{figure*}
We begin with the presentation of the profiles obtained during the two epochs.  These are then used to derive the electron density and mass of the ejecta and we present a unifying model for the evolution of the profiles.  The strongest lines in the optical spectrum were the Balmer and nebular forbidden lines of [O III]. 
%%%%%%%%%
% check the Fig. 1 spectra, these are the He sequence with Balmer
%%%%%%%%%%%%%

%Note that the [N II] 6548, 6883\AA\ lines are present in the wing of H$\alpha$ and alter the high velocity parts of the profile.   The high resolution is needed to show the shoulders on the profile.
\subsection{Line profiles in the two epochs}

The H$\alpha$ and H$\beta$ lines profiles are shown in Fig. 1.  The blending of H$\alpha$ with the [N II] doublet 6548, 6583\AA\ is more clearly seen in the epoch 2 spectrum because of the reduced flux relative to epoch 1.  In contrast, the H$\beta$ line was unblended at both epochs and we used this as the standard of comparison (see later discussion of the [O III] and [N II] lines).  

He I 5875, 6678, 7065\AA\ were detected in epoch 1 but none were present in the epoch 2 spectrum (Fig. 1).   The He II 4686\AA\ profile was strongly blended with the N III complex at 4640\AA\ but a comparison with He I 5875\AA\ shows that it displayed the same profile (Fig. 2) .   The He II 5411\AA\ line was not present in either spectrum. The 4686\AA\ line was narrow and present on the epoch 2 spectrum.  The high velocity wings, with $|v_{rad}|>1000$ km s$^{-1}$ were very much reduced relative to the central emission.  As we will discuss in sec. 3.3, this was a general feature of the profile development.  The 4640\AA\ feature was still present at epoch 2 but its relative peak flux was reduced by a factor of 4 relative to He II 4686\AA.

The optical G430L STIS spectrum  was used to calibrate the epoch 1 NOT observation by convolving the spectrum following median filtering to the STIS resolution (about 3\AA)  and using a third order fit to the strongest lines from 3300-5700\AA.  The fluxed spectrum was dereddened using E(B-V)=0.5 (Shore et al. 2012) (see Fig. 2).  Despite the strong UV blend with O III] 1667\AA\ and the optical with the 4640\AA\ complex, the clear similarity of the two He II profiles is evident.  The dereddened fluxes were F(He II 1640) = $(5.0\pm 0.1)\times 10^{-10}$ erg s$^{-1}$cm$^{-2}$, F(He II 4686)=$(3.0\pm 0.2)\times 10^{-11}$ in the same units, so the flux ratio was $\approx$ 16.5$\pm$0.2.   The epoch 2 He II 1640\AA| line had the same narrow core as the optical forbidden lines and a low extended pedestal with $|v_{rad,max}| = 2000$ km s$^{-1}$.  The measured flux was (6.7$\pm$0.1)$\times10^{-13}$ erg s$^{-1}$cm$^{-2}$, the dereddened flux was 2.3$\times 10^{-11}$.    For the epoch 2 data, the dereddened fluxes for He II 4686 \AA\ was (1.9$\pm$0.2)$\times$10$^{-12}$ erg s$^{-1}$cm$^{-2}$.  We note that the decreased flux is  consistent with a constant mass ejecta.    A more complete analysis, including a discussion of the He/H ratio (an open question for recurrent novae)  will be presented in the next paper (Schwarz et al 2012, in preparation).  

\begin{figure}
   \centering
   \includegraphics[width=8cm]{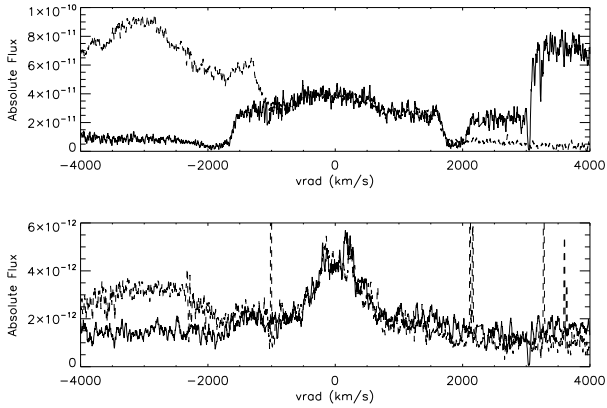}
   \caption{Comparison of He II 1640 \AA\ (solid) with He II 4686 \AA\ (dot-dash) from epoch 1 (top) and epoch 2 (bottom).  All spectra were dereddened with E(B-V)=0.5.  For display purposes, the optical NOT spectrum is multiplied by a factor of 40 with a boxcar smoothing to 11 points. }
    \end{figure}

\begin{figure}
   \centering
   \includegraphics[width=9cm]{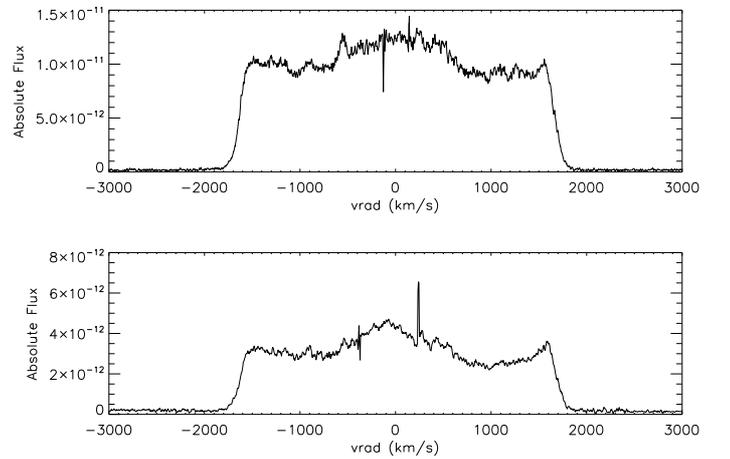}
   \caption{Comparison of the N IV] 1486 \AA\ (top) and C III] 1909 \AA\ (bottom) profiles from the STIS epoch 1 spectrum.  The C III] line was not covered  in the epoch 2 STIS observation. }
 \end{figure}

%FIG 5
\begin{figure}
   \centering
   \includegraphics[width=9cm]{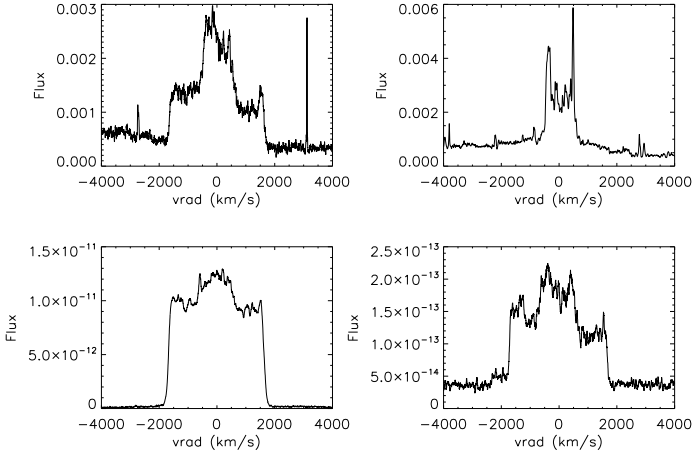}
   \caption{A comparison of the simultaneous nitrogen line profiles from epoch 1 (left) and epoch 2 (right).  Top: [N II] 5755 \AA, bottom: N IV] 1486 \AA.  Note, in particular, the change in the core to wing ratio (see text for further discussion).}
\end{figure}

As we show in Fig. 3, the epoch 1 isoelectronic ultraviolet resonance intercombination lines of C III] 1909 \AA\ and N IV] 1486 \AA\ had the same line profiles, essentially the same as H$\beta$.   The N III]1751\AA, N IV] 1486 \AA, and N V 1238, 1242 \AA\  lines in the epoch 1 STIS spectrum showed the variation of the ionization in the ejecta, see Fig. 5.   The N III feature  is, however, a complex blend of permitted  ($^2$P-$^2$D$^o$) and intercombination ($^2$P$^o$-$^4$P , resonance) lines.  Assuming that the individual profiles are the same as the C III] 1909\AA\ line, the Einstein-A weighted combination profile produces a very close match to the observations.  This is the same result we found for the H$\alpha$ plus [N II] optical lines and highlights that we are tracing not only an ionization but also density structure.  The N V UV resonance doublet is, however, quite different and {\it cannot} be reproduced by any naturally weighted combination of individual components.  There appears to be a real difference on the blueshifted wing, the ionization is lower on the approaching side of the line.   There was another important difference.   The N V doublet showed a strong, relatively narrow peak at around 1500 km s$^{-1}$ in both spectra while that component vanished on C IV 1550 \AA\ in the epoch 2 spectrum (Fig. 5).  We compare these to the Si IV/S IV] blend at 1402 \AA\ for which we expect the dominant contributor to be from O$^{+3}$.    The lower ionization lines for nitrogen are shown in Fig. 4.

  \begin{figure}
   \centering
   \includegraphics[width=9cm]{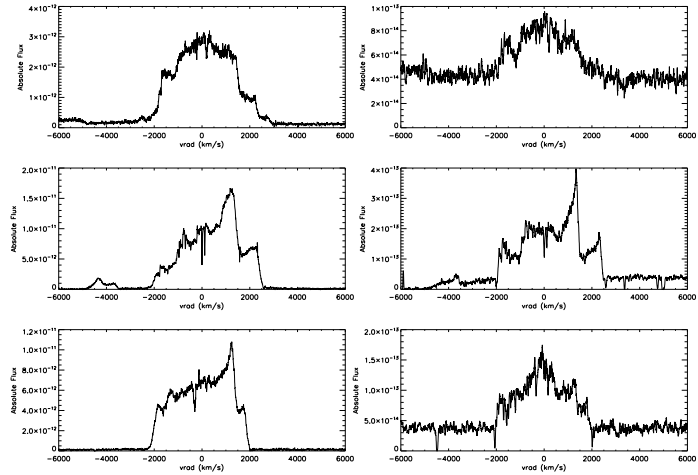}
   \caption{The ultraviolet resonance multiplets Si IV/S IV] 1402 AA\ (top), N V 1240 \AA\ (middle) and C IV 1550 \AA\ (bottom)  for epoch 1 (left) and epoch 2 (right).  The interstellar absorption components are visible on the resonance lines.  For comparison the mean wavelength of the blend has been used.  Note the strong change in the C IV line at about +1300 km s$^{-1}$, but note that the C IV and N V are blends with large velocity separation between the components and the variations cannot be uniquely assigned.   See text for details.}
    \end{figure}

The [O III] 4363, 4959, 5007\AA\  nebular lines were present and strong in both spectra, as shown in Fig. 6, but displayed different profiles than the He II.  The three lines show the similar profiles to H$\beta$ for each epoch, as a contrast between Figs. 1 and 6 shows.  The  lines closely resembled the optical N$^+$ transitions that are their isoelectronic analogs.  In the next section, we will discuss the use of these lines to determine the electron temperature and density in the two epochs.

 \begin{figure}
   \centering
   \includegraphics[width=9.3cm,height=10cm]{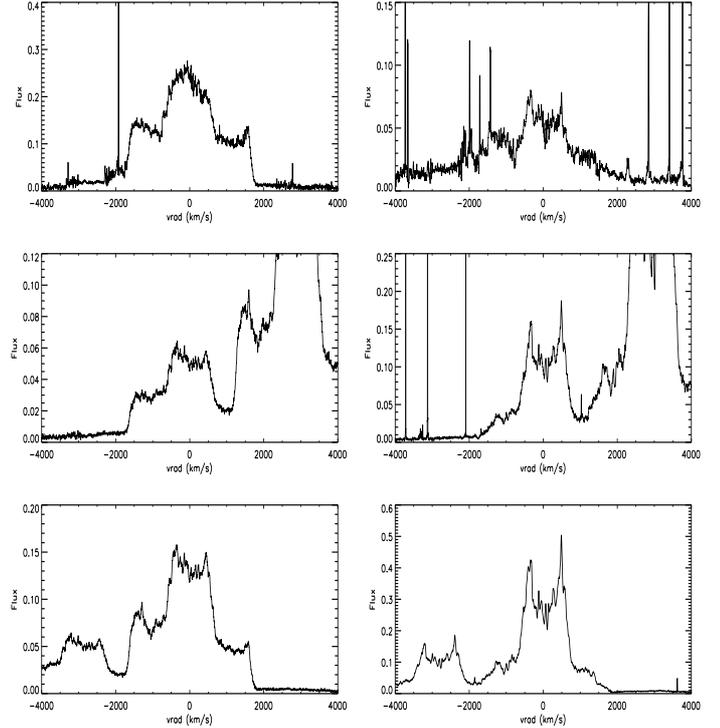}
   \caption{The [O III] line profiles, in velocity, from  epoch 1 (left) and epoch 2 (right).  From the top, $\lambda$ 4363 \AA\, $\lambda$ 4959 \AA\, and $\lambda$ 5007 \AA.  The fluxes have not been corrected for reddening and no absolute calibration was applied to the epoch 1 data for display purposes.  See text for discussion.}
    \end{figure}        

The high ionization [Ca V] 5309, 6086 \AA\ lines were present in the epoch 1 spectrum and the 6086 \AA\  line was also present  at epoch 2 but the 5309 \AA\ line was much weaker in this noisy part of the spectrum.   The 6086 \AA\ profile was similar to the He II transitions, while the 5309 \AA\  line was asymmetric but with a FWZI about the same as the other emission lines.\footnote{Although an alternate identification for the 6086 \AA\ line is [Fe VII],  other lines of that ion are not detected at either epoch.}  The two lines are shown in Fig. 7.

  \begin{figure}
   \centering
   \includegraphics[width=9cm]{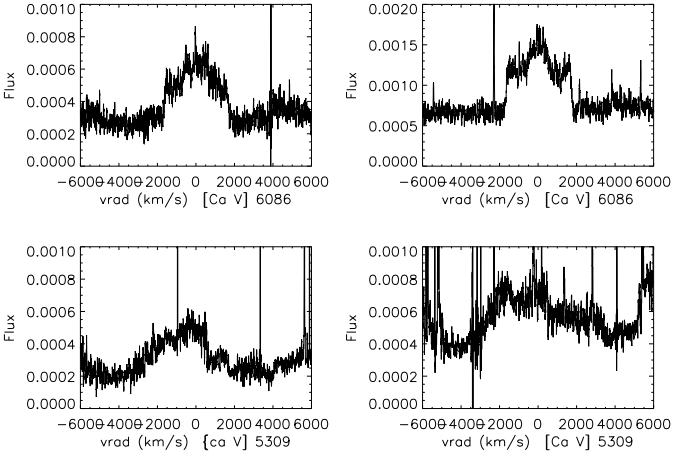}
   \caption{The [Ca V] 6086\AA\ (top) and 5309 \AA\ line profiles for epoch 1 (left) and epoch 2 (right).   }
    \end{figure}

\subsection{Lyman $\alpha$ analysis and neutral hydrogen column density}

The Ly$\alpha$ line was observed at both epochs.  It is heavily extinguished by interstellar absorption but there appears to be a second emission line in the redward wing.  To attempt an identification and to determine the extinction, we have modeled the intrinsic Ly$\alpha$ profile using the epoch 1 H$\alpha$ line as a proxy together with a velocity shifted weighted copy.  The derived wavelength for the presumed blend, 1318\AA, is roughly coincident with the intercombination resonance line O V] 1318.34\AA\ as shown in Fig. 8.\footnote{We note that the epoch 1 spectrum also showed O V 1371, which was absent in the epoch 2 observation.  We postpone further discussion to the next paper.}   To our knowledge, this is the first identification of this transition in a recurrent nova (it has recently been identified by Young et al. (2011) in symbiotic star spectra) .  It was absent in the epoch 2 spectrum.  The interstellar absorption gives a column density $N_H = (3\pm0.5)\times 10^{21}$cm$^{-2}$ that is consistent with the reddening, E(B-V)=0.5, proposed in Paper I.   Although the continuum between 1800 and 2400 \AA\ is weak, the strength of the 2175 \AA\ feature is consistent with this reddening.  An important feature of this simulation is the indirect verification of the transparency of the ejecta, since the model assumes the same profile for Ly$\alpha$ as H$\alpha$.  The upper limit for the emission in the epoch 2 spectrum is compatible with the overall fading of the ultraviolet emission lines.  We also note that the optical [Ca V] lines had also decreased substantially by epoch 2, consistent with this change in a high ionization species.

\begin{figure}
   \centering
   \includegraphics[width=9cm]{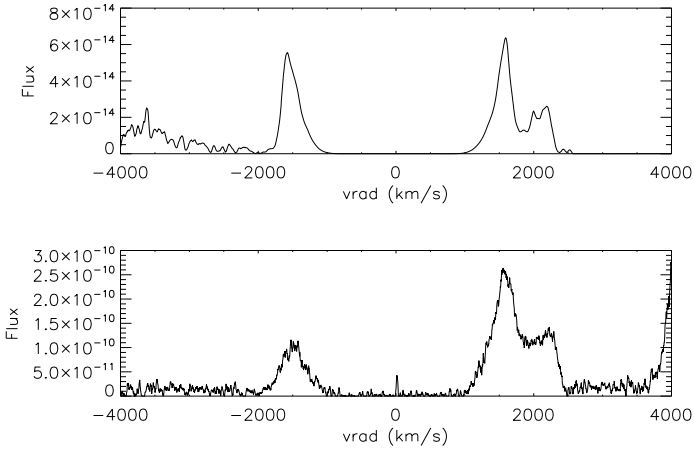}
   \caption{top: Epoch 1 Ly$\alpha$ profile model using H$\alpha$ and O V] 1318.3\AA.  Bottom: Epoch 1 Ly$\alpha$ STIS observation.}
    \end{figure}

\subsection{Modeling the structure of the ejecta}

Understanding the properties of the resolved ejecta of T Pyx has been a daunting task. The ejecta present an extremely fragmented appearance in almost every emission line filter.  The HST images show a nearly symmetric but completely filamented ring (e.g. Shara et al. 1997).   This confusion has been dramatically removed  by the infrared continuum and line spectro-interferometric observations during the current outburst by Chesneau et al. (2011) using infrared CHARA and VLTI .  They find that 
%during the peak of the light curve, before day 20, the expansion was $\le$300 km s$^{-1}$ while 
after day 28 the AMBER K-band and PIONIER-H band visibilities are consistent with a two-component bipolar plus central point source model with very low inclination.   In light of the uncertainties in the orbital inclination from Uthas et al. (2010), we set our constraints based on the interferometry.

Models of nova ejecta have long included strong departures from simple expanding spheres.  In her fundamental monograph, Payne-Gaposchkin (1957) highlighted examples of the variations of the line structures, invoking rings and shells to reproduce the individual components.  Hutchings (1970, 1972) included polar cones and rings when modeling the profiles in HR Del 1967.    The (ultimately) spatially resolved ejecta of this nova have also been studied by Solf (1983) who combined an equatorial ring with conical polar ejecta to reproduce the line profiles.   This was elaborated by Hillwig (2001), using groundbased optical multifiber spectroscopy and HST/WFPC2 imaging.  The HST H$\alpha$ and [O III] images show strikingly different structures,  H$\alpha$ being more extensive and thicker while the forbidden lines show thin-walled polar cones of the sort discussed by Solf and also well known from planetary nebulae.  

We used a Monte Carlo procedure to model the optically thin nebular spectra.  The ejecta geometries were assumed to be 
 either spherical or axisymmetric spheroids (prolate or oblate).  The free parameters of the models were the opening conical (bipolar) angles (an inner angle $\theta_1 \ge$ 0, and an outer angle $\theta_2 \le$ 90$^{\circ}$),  the inner radial extent, $\Delta R/R_{max}$, relative to the outer radius $R_{\max}$ that is given by the maximum velocity $v_{max}$, and an inclination $i$ to the line of sight (for the spheroidal and conical
models).  The position angle in the plane of the sky was fixed at o$^o$.  The velocity was assumed to be linearly dependent on the radius (a ballistic or so-called ``Hubble'' flow) with the density varying as $r^{-3}$ (constant mass for the ejecta) and a power law
density dependence for the emissivity, $\rho^{n_d}$.  In general, we assumed $n_d=2$ for  recombination lines.  This method is an extension of the procedure used in our previous studies, e.g.  V1974 Cyg (Shore et al. 1993), V382 Vel (Shore et al. 2003), and V1186 Sco (Schwarz et al. 2007).    A sample of line model profiles is shown in Fig. 9 and examples of the resulting images and a representative line profile are shown in Fig. 10.   Gill \& O'Brien (1999) presented an ensemble of computed spectra and images using ring geometries and various ellipticities for the ejecta that show many of the features we will describe.  In this our efforts can be seen as an extension of the existing models but also a unification of the phenomenology.  Harman \& O'Brien (2003) modeled HR Del with similar assumptions.   Thus, while such structures -- polar ejection, rings, and shells -- have been frequently suggested in the literature since the start of modern nova studies  (e.g. Gill \& O'Brien 1993), the novelty here is the ease of modeling provided by the Monte Carlo method and the specific dynamics.

%FIG 10
\begin{figure}
   \centering
   \includegraphics[width=8cm]{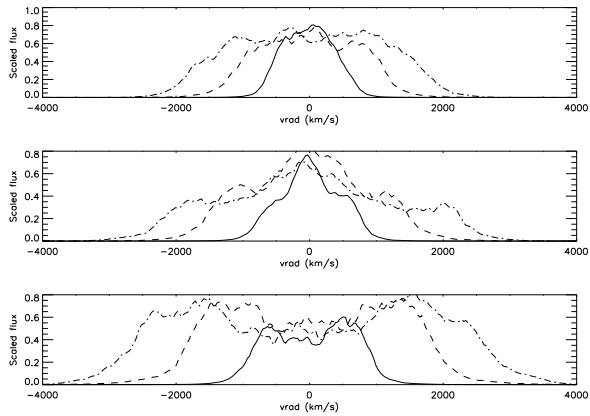}
   \caption{Examples of simulated line profiles for $\theta_o = 5^o$ and $\theta_i=30^o$.  Solid: $\Delta R/R$=0.4; dash: $\Delta R/R$=0.6; dot-dash: $\Delta R/R$=0.8.  Top panel: $i=15^o$; middle panel: $i=30^o$; bottom panel: $i=60^o$.  The maximum velocity was 4000 km s$^{-1}$.       This shows the range of profiles expected for a sample of recurrent novae with similar structure to T Pyx (see text for details).}
    \end{figure}

At fixed inclination, the full width at zero intensity (FWZI) and full width at half-maximum (FWHM) of the profiles depend critically on $\Delta R/R$.  In the earliest spectra, taken with the NOT during the fireball and
optically thick stages (Shore et al. 2012) the maximum velocity of the P
Cyg troughs on the Balmer and He I optical lines was about 4000 km
s$^{-1}$.  Using this as the maximum velocity for the ejecta, $0.4 \le
\Delta R/R \le 0.5$, independent of the inclination or the exponent of the
intensity power law.   The structure of the line  peak, in contrast, has a mixed
dependence on the opening angle of a polar ejection, the inner polar angle,
and the inclination.  Since the Balmer, [N II], and [O III] lines all
display the same profiles in the spectra we discuss here, it
appears that there is no particular difference between lines formed by
radiative excitation (e.g.the forbidden lines) and those from recombination
(Balmer, He II).

\begin{figure*}
   \centering
   \includegraphics[width=16cm]{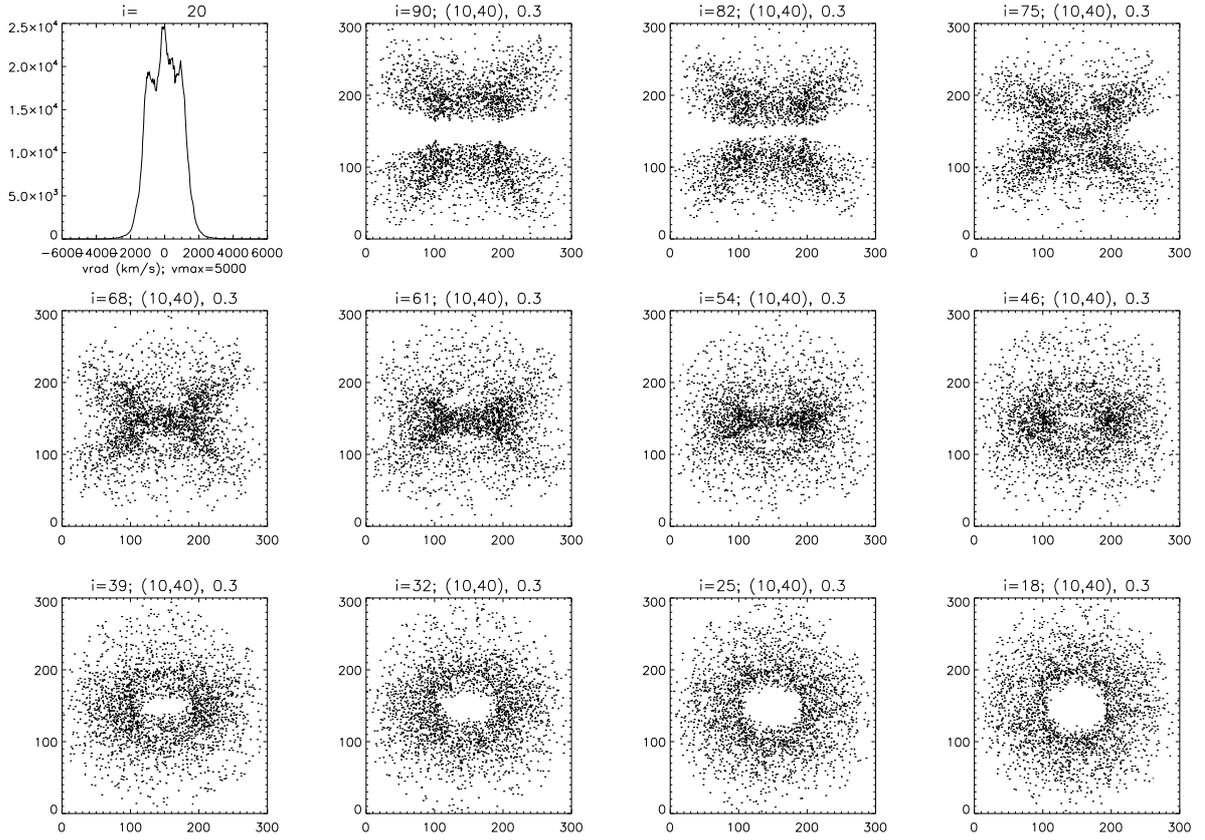}
   \caption{Example of a line profile and synthetic images, assuming that the lines are formed by recombination and with ballistic velocity law, constant mass bipolar ejecta  for various inclinations (as noted on each frame).  All models  assumed $v_{\rm max}=5000$ km s$^{-1}$, with the outer and inner angles (10,40) and $\Delta R/R =0.3$.  The profile shown in the upper left is for $i=20^o$.}
    \end{figure*}

An important clue regarding the geometry comes from the comparison of the high velocity portions of the line profiles in the two epochs.  These decrease in intensity while the core to wing ratio increases.    The simulations match, shown in Fig. 11 for each epoch, the line profiles shown in Fig. 1 for the Balmer and He I lines and Fig. 4 for the nitrogen transitions.  The He II 1640, 4686\AA\ lines have much more reduced emission on the ``shoulders'' of the line in epoch 2 than the forbidden lines.  This can be reproduced by assuming either a smaller inner fractional radius and/or changing the
radial dependence of the intensity power law to mimic different ionizations without a change in the overall geometry of the ejecta.   Figure 12 shows the two UV line profiles observed in both epochs, N IV] 1486\AA\ and He II 1640\AA.  As we described, the He II UV and optical 4686\AA\ profiles are identical.  The He$^+$ region we find is more extensive in the ejecta, signaled by the narrower line profile, but the cone parameters are rather similar.   The UV resonance intercombination lines all displayed the same profiles so we consider the match obtained here to be generic of that line forming portion of the ejecta.  The major anomaly remains the N V 1240 \AA\ and C IV 1550 \AA\ doublets.  In these doublets there seems to be a blend that is not modeled by the respective resonance lines alone.  The success of the modeling for other complex lines, e.g. N III] 1751 \AA, highlights this peculiarity.  The relative changes in the two sides of the line appear more strongly linked to the 1548 \AA\ component but this remains a puzzle.  
%FIG. 11

\begin{figure}
   \centering
   \includegraphics[width=9cm]{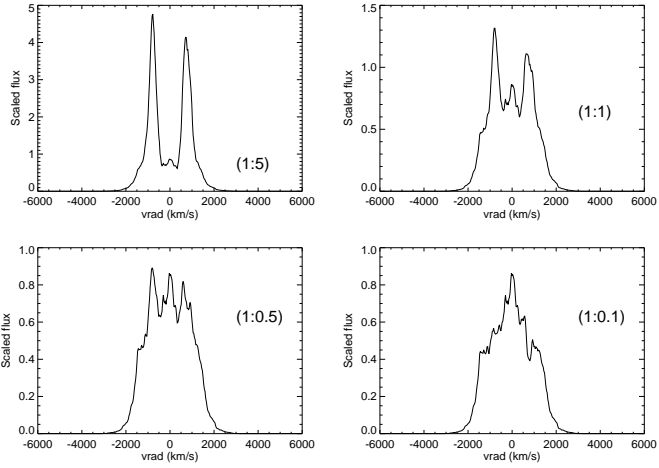}
   \caption{Composite bipolar lobe line profiles.  All simulations used $v_{max}|=4000$ km s$^{-1}$, $i=20^o$, and ballistic, constant mass ejecta.  We use the notation, here and in Fig. 12,  [$\theta_0,\theta_1,\Delta R/R$,$i$] for the model parameters.    The broad component (consistent with the epoch 1 data) has the parameters [5,30,0.5,20], and is combined with a second consistent with the  [O III] profile (see fig. 7) with [50,90,0.2].  The relative weights of the two contributing profiles are indicated in each frame.  See text for details.}
       \end{figure}

Acceptable fits were obtained for a range of inclinations, depending on the assumed
values for the angles.  There is one very strong constraint: inclinations
as low as 20$^{\circ}$ require an oblate symmetry with polar plumes for
almost any angle.  In general, the profiles cannot be reproduced with
narrow opening angles, $\theta_1 >$ 20$^{\circ}$, 
60$^{\circ} < \theta_2 \le$ 90$^{\circ}$, 
because of the distinct separation of the polar ejecta in
velocity.  In contrast,  nearly spherical ejecta -- while lacking the
strong central peak that is produced most easily at high inclination by the
superposition of the two projected expansion velocities of the polar cones
-- give acceptable agreement without the central peak - shoulder structure.
None of these details yield any particular value for the filling fraction,
the filamentary/knotty structure of the ejecta is a natural consequence of
numerical noise in the simulation.  Any velocity law analogous to that from
radiatively driven outflows from massive stars is ruled out by the model profiles for all
inclinations.
%and good agreement with the dispersion in
%intensity with velocity is obtained for relatively low numbers of emitting
%volumes.   %FIG 11 -- REDO THI S
 \begin{figure}
   \centering
   \includegraphics[width=9cm]{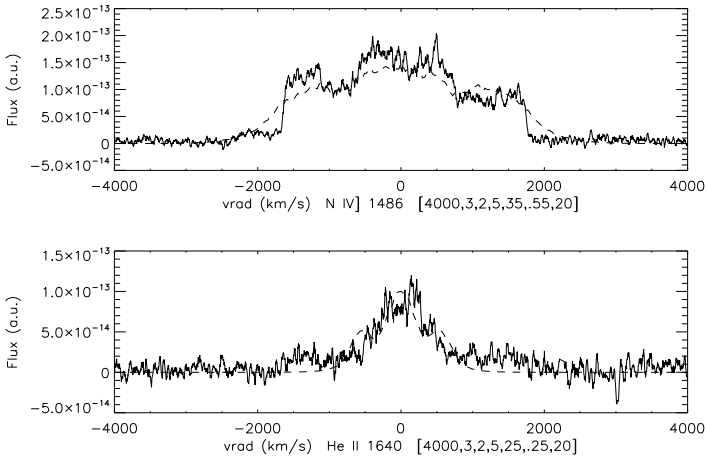}
   \caption{N IV] 1486\AA\ and He II 1640\AA\ from the STIS 2012 Mar. spectrum with model line profiles.  The parameters for the model are indicated for $v_{\rm max}$=4000 km s$^{-1}$  and i=20$^o$.}
    \end{figure}

\subsection{Electron density and mass of the ejecta}

For epoch 1, the [O III] 4363, 4959, 5007 \AA\ lines were strong and not severely blended with other lines.  We scaled the H$\beta$ profile to H$\gamma$ to remove the small ($\sim$10\%) residual in the [O III] 4363 \AA\ profile and calculated the electron density as a function of radial velocity by taking the ratio of the continuum subtracted resolved line profiles.  
%The result is shown in Fig. --.  
The derived value depends on the electron temperature, T$_e$.  For the range $1\le (T_e/10^4K) \le 4$ we obtain $5.4 \ge (n_e/10^7\ cm^{-3}) \ge  0.9$ that is almost constant across the profile (Osterbrock \& Ferland 2006).   The [N II] 6583\AA\ line was very weak but present in the high velocity wing of H$\alpha$.  Since the NOT spectrum was obtained after the transition to transparency, we used the scaled H$\beta$ profile in an attempt to disentangle H$\alpha$ and [N II] 6548, 6583\AA\ to have an independent estimate of the electron density (see Fig. 13).   The residual emission was very weak but the ratio $F(5755)/F(6583) \le (0.5\pm0.1)$, corrected for E(B-V)=0.5, which for the same range of T$_e$ corresponds to $4.9 \ge (n_e/10^6\ cm^{-3}) \ge 1.4$.  No single combination of $n_e$ and T$_e$ yields both diagnostic ratios.  In contrast, since for epoch 2  the  [N II]/H$\alpha$ ratio had significantly increased, we were able to obtain a single set of physical properties.  The observed, dereddened [O III] ratio was $F(4949+5007)/F(4363)=8.4\pm0.1$.   Using the scaled H$\beta$ profile as before, we recovered the [N II] 6583 \AA\ line (see Fig. 12).  The 5755\AA\ and  the recovered 6583\AA\ profiles are identical.  The flux ratio, $F(6548+6583)/F(5755)=1.2\pm0.1$, is approximately constant for $|v_{\rm rad}| < 1000$ km s$^{-1}$ with the exception of the narrow feature at around +600 km s$^{-1}$.  A single pair of parameters results from the two epoch 2 diagnostics, T$_e=3.9\times 10^4$K and $n_e = 5.4\times 10^5$cm$^{-3}$, see Fig. 14.  The similarity of all profiles at this second epoch supports the picture of nearly isothermal, transparent ejecta (see Fig. 13).   As a further check of the assumption of a constant mass for the ejecta for which $n_e \sim t^{-3}$, the density should decrease by about a factor of 10 between the two epochs (Day 170 and Day 350 after discovery).  This roughly agrees with the He II 1640 \AA\ and 4686 \AA\ flux variations.   

Using the electron density for epoch 2, and assuming v$_{max} = 4000$ km s$^{-1}$ (based on the earliest optically thick spectra, see Paper I), the mass derived for the ejecta is approximately $5\times$10$^{-5}f$ M$_\odot$ independent of the distance.  The solid angle of the ejecta obtained from modeling the profiles is independent of $f$, which instead measures the fragmentation of the expelled gas.  From the complexity of the narrow absorption features on the Balmer lines (see Paper I), we expect that $f$ may be much less than unity so this mass is clearly an upper limit.    An estimate of the filling factor can be derived from the Balmer line fluxes from the epoch 1 data.  The measured H$\beta$ flux from the G430L STIS spectrum from epoch 1 was $1.5\times 10^{-11}$ erg s$^{-1}$cm$^{-2}$.  Corrected for E(B-V)=0.5, this corresponds to $F(\beta)=7.7\times 10^{-11}$ erg s$^{-1}$cm$^{-2}$.  Taking a distance  $\ge$3.5 kpc gives a luminosity $L(H\beta) \ge 1.2\times 10^{35}$ erg s$^{-1}$.  Using the lower distance and assuming case B recombination for  T$_e$=$10^4$K yields an emission measure of $n_{e,7}^2V \approx 1.2 \times 10^{47}$ cm$^{-3}$ where $n_{e,7}$ is the electron density in units of $10^7$cm$^{-3}$ for source volume $V$.   Taking $n_{e,7} \approx 3$ for Day 170 gives an estimate of $f \approx 0.03$.\footnote{The uncertainty is difficult to constrain since we are choosing only the lower limit on the distance but the extinction uncertainty is about 10\% for the luminosity.}  This is in the range of typical values for the filling factor (see e.g. Shore (2008) and references therein).  The implied ejecta mass, M${ej} \approx 2\times 10^{-6}$M$_\odot$ is  similar to that obtained for other RNe.  In contrast, Schaefer et al. (2010) derived a mass for the matter that they identify with a single classical nova-like outburst of 1866, M$_{ej}\sim 3\times 10^{-5}$M$_\odot$.   Selvelli et al. (2009) and, more recently, Evans et al. (2012) have also argued that the mass of the accumulated flotsam of previous ejections may be considerably greater, perhaps a factor of order 100, than that of a single event.   Evans et al. (2012) based this on the infrared light curve of the 2011 event that they explain as a light echo from the previously ejected material.\footnote{If our estimated mass is typical of an explosion of T Pyx,  the circumstellar matter could be a  cumulative production of successive outbursts.  This implies an activity period of several millennia, at least.} 

 \begin{figure}
   \centering
   \includegraphics[width=9.5cm]{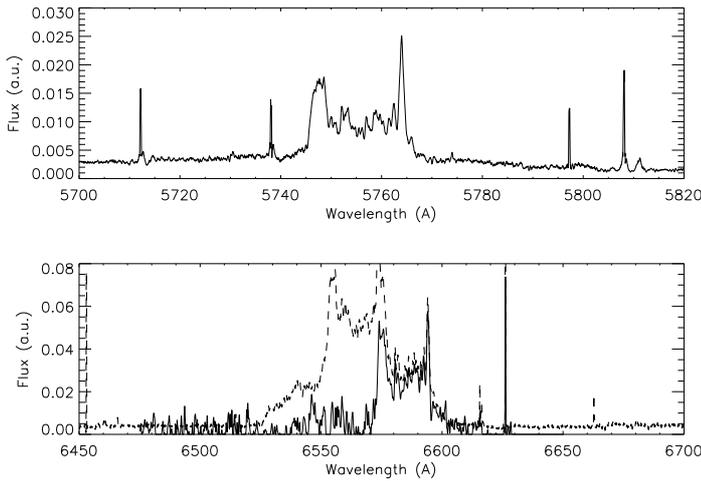}
   \caption{Optical N II profiles, 2012 Apr., used for the electron density analysis.  Top: [N II] 5755\AA; bottom: [N II] 6583\AA\ (solid), composite profile with H$\alpha$ before decomposition (dash).  If present, [N II] 6548\AA\ is extremely weak.  See text for details.}
       \end{figure}        
        
 \begin{figure}
   \centering
   \includegraphics[width=9cm]{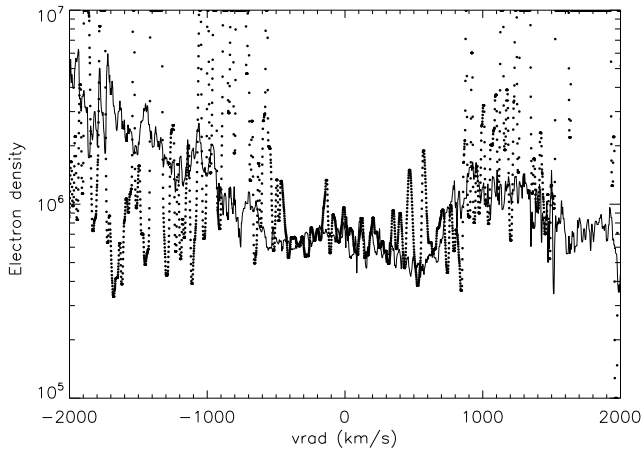}
     \caption{Comparison of [N II] (dots) and [O III] (solid line) electron densities from epoch 2.  The wings beyond $|v_{rad}=1000$ km s$^{-1}$ are comparatively weaker in [N II] than [O III], hence the much higher noise.}
    \end{figure}

\subsection{Summary of the modeling}

The observed line profiles during the optically thin stages of the expansion of T Pyx can be explained by a  unified model in which the ejecta have an axial (bipolar) symmetry with the expansion directed at low inclination to the line of sight.  This agrees  with the interferometric imaging {\bf (and perhaps also with the infrared emission described by Evans et al. (2012))}.  The relative contributions of the inner and outer portions of the polar lobes change  systematically in time.  The observed line profiles for all ions, except N V 1240 and C IV 1550 -- conform to this  scheme.  The differences between RNe is then mainly consequences of the differences in inclination of the ejecta axis to the line of sight and the thickness of the line forming region relative to the maximum velocity.  The profiles are interpretable using ballistic ejection, resulting in a linear velocity law, and constant mass.  While not definitively excluded, there is no need to invoke either continuing outflows, such as super-Eddington stellar winds, or multiple individual ejection events to obtain the structures in the observed line profiles during the nebular phase.  By extension, this may apply to other classical novae for which similar geometries  and variations have been reported in the literature.  For any inclination, for $\Delta R/R \ge 0.2$, an outer angle of $\theta_o > 20^o$ is not able to reproduce the profiles for any  inner angle $\theta_i$.  The profiles are always two well separated, although not symmetric, peaks.  There is general agreement in the literature that the inclination must be quite low but there is still a considerable range, from $\sim$6$^o$ to10$^o$ (favored by optical interferometric imaging and orbital solution) to $\le$30$^o$ (Selvelli et al. 2009), the FWZI of the profile is uniquely determined by the ejecta thickness.

\section{Discussion}

The simplest way to understand the line profiles is with polar cones rather than either rings or multiple shells but these are the zones in which the emission lines are formed.  They may not, however, completely delineate the mass distribution.  That requires detailed photoionization modeling (e.g. Ercole et al. (2003)).   But on the basis of these simulations we suggest that the recurrent novae\footnote{We exclude in this discussion the symbiotic-like recurrent novae, e.g. RS Oph, T CrB, and V407 Cyg.  In such systems the effects of the circumstellar wind environment dominate the phenomenology.  But it is possible that these too are ejected in a predominantly bipolar geometry at the WD.}, and by extension perhaps also classical novae, are unified by the structure observed in the profiles of the compact sources.  Observations of the known RNe, e.g. CI Aql (Iijima 2012), and the RN candidate KT Eri agree morphologically with those obtained in this study of T Pyx.  The main difference, aside from the photometry, is the early presence in the most compact systems, those with orbital periods less than about 0.5 days, of an optically thick stage that closely resembles classical novae during the rise to maximum visible light.

  \begin{figure}
   \centering
   \includegraphics[width=9cm]{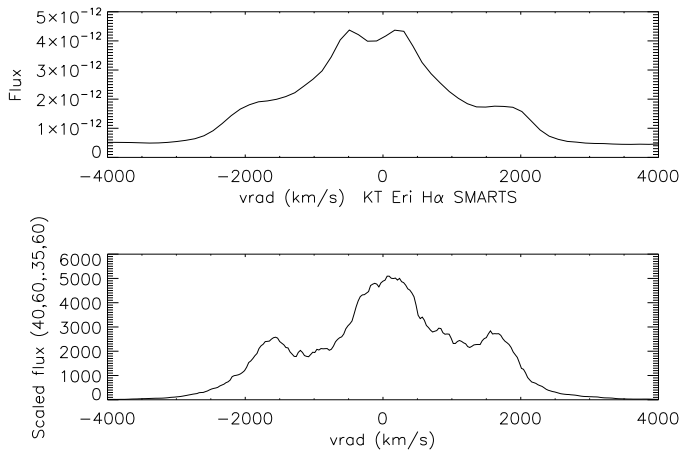}
   \caption{Comparison of the H$\alpha$ line for KT Eri from 2009 Dec. 12 (Day 18),  a SMARTS high resolution (3\AA) spectrum (top) with a model profile with angles (40,60) degrees, $\Delta R/R=0.35$ and $i=60^o$.  The range in $i$, $\Delta R/R$, and angles is about 10\%. }
    \end{figure}

As an extension of the technique, we have applied the same geometry to several other, well observed, recurrent novae.  For KT Eri, the line profiles during the nebular stage obtained by the SMARTS\footnote{URL: http://www.astro.sunysb.edu/fwalter/SMARTS/NovaAtlas/ nova survey}  (Walter et al. 2012) are almost identical to the T Pyx spectra at a similar stage in the light curve, although with a higher maximum velocity.  An example of applying the modeling technique is shown in Fig. 15.  This suggests that the ejecta thickness may have been slightly different between the two but the profiles are similar.  The late time STIS/HST spectrum of U Sco 1999 at H$\alpha$ (o5fq08010, G750M) and the spectra shown in Iijima (2002, figs. 2 (Day 1) and 7 (Day 17)) are also quite similar once care has been taken to remove the contaminating blend with the [N II] lines.
%\section{Concluding remarks and speculations}

The similarity of the T Pyx line profiles during the nebular stage to  RNe of the U Sco type leaves the early {\bf long duration}  recombination wave event modeled in Shore et al. (2011) paradoxical.  The ejecta mass is  small, $<$10$^{-5}$ M$_\odot$, yet the spectral variations in the first two months following discovery displayed the recombination wave (Shore et al. 2011) that is expected only in optically thick gas with column densities greater than $10^{23}$cm$^{-2}$.  In fact, the prolonged maximum indicates a very significant opacity in the ultraviolet, more like a classical nova for which the ejecta masses are  a factor of 10 greater or even more.  {\bf The low mass derived here for the T Pyx ejecta, based on the optically thin stage, is comparable to those derived for other recurrent novae.   However, there is a clue.   There is a very prolonged optical maximum -- corresponding to an extended optically thick stage  lasting nearly two months -- for T Pyx, IM Nor, and CI Aql.}    For V394 CrA, U Sco, and the symbiotic-like recurrent novae, the early optical light curve that is dominated by the ultraviolet opacity changes in the ejecta show nothing so extended in duration.  The interferometric images indicate a similar structure to the ejecta for T Pyx as many other novae, even classical types such as HR Del.  The filling factors of the ionized regions are quite low.  From the comparison of the line profiles of individual ionization stages we have shown that the ejecta can best be explained as bipolar structures of moderate thickness and in ballistic expansion.   All indications point to a low inclination for the  T Pyx orbital plane, of order 10$^o$, while U Sco is, in contrast, an eclipsing system.   The main difference seems to be the orbital period: {\it only} the compact systems with P$_{\rm orb}$ less than about 0.6 days show the extended optically bright phase, based on the data discussed in Schaefer (2010).  {\bf We suggest that the extended opaque phase could be due to the formation of a cool common envelope after the explosion, causing a recombination wave to move outward through the ejecta that also extinguishes the XR emission.  It could then slowly clear as the WD settles into a stage of quasi-static nuclear burning and develops a supersoft source.}   We emphasize that this is only a hypothesis at present, current models do not permit the computation of fully three dimensional evolution of the system  through this phase.  {\bf It could, however, be related to the axial symmetry of the ejecta, a question yet to be explored (we thank the referee for this comment).}

Finally, a particularly intriguing result we are investigating  is that the Balmer and metal line absorption systems described in Paper I correspond to the blueshifted  peaks in the emission line profiles from months later.  The last NOT spectrum from Paper I, obtained on 2011 May 30 still during the optically thick stage, shows the absorption on the Balmer lines confined to within the same velocity range as the wing of the optically thin epoch 1 profiles and the emission peak and structure on the redward portion of the Balmer lines showing similar characteristics to the later emission (see Fig. 15).  This is, in our view, another strong argument that the absorption, and the attendant ``accelerations'' are produced by the recombination wave within the expanding ejecta and not something arising in the circumsystem medium.  The maximum absorption velocity observed before 2011 May. was -2000 km s$^{-1}$, virtually identical to the limits of the emission lines, and structure appearing in the redshifted wings also mimics the later observations.  This further implies that the material was confined even at the earliest stages to the narrow cone (and possibly ring) found from interferometry and our modeling.  However, there is no possibility that with such a small mass, and so narrowly confined toward the periphery, that the ejecta alone could have produced the required column densities for the absorption/fluorescence spectrum and the recombination event.   

  \begin{figure}
   \centering
   \includegraphics[width=9cm]{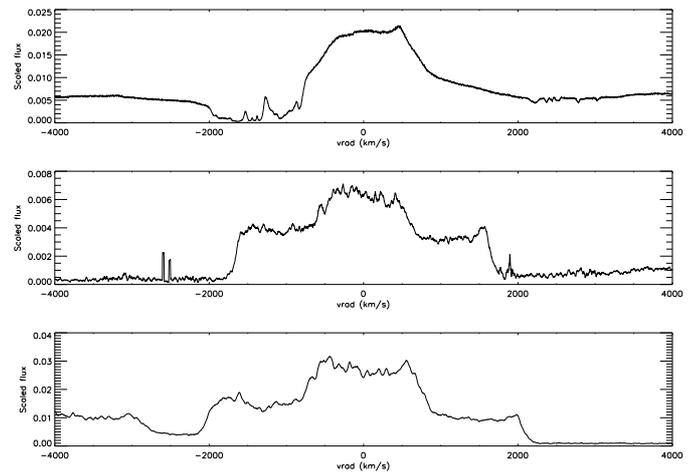}
   \caption{Comparison of H$\beta$ from 2011 May 30 (top), H$\beta$ 2011 Oct. (epoch1, middle), and [O III] 5007\AA\ from epoch1, bottom).  Notice the similarity of the nascent emission features in the line peak and redward portion and those in the later optically thin spectra.  See text for details. }
    \end{figure}
 
        \begin{acknowledgements}

We thank the referee, G. C.  Anupama, for thought provoking questions, and  Mike Bode, Jordi Casanova, Nye Evans, Jordi Jos\'e, Tim O'Brien, Brian Warner, and Bob Williams  for discussions and correspondence.  We also thank  Olivier Chesneau for his comments on the draft.  We have made extensive use of the Astrophysics Data System (ADS), SIMBAD (CDS), and the Barbara A. Mikulski MAST archive (STScI) during this work.   The STIS spectra were obtained in program GO 12200 and GO/DDT 12799.  The NOT observations were obtained in Fast Track proposal 43-403.
  
\end{acknowledgements}

    \end{document}